\documentclass[twocolumn]{aa}  

\usepackage{graphicx}
\usepackage{txfonts}
\usepackage{hyperref,ulem}
\usepackage{color}
\usepackage{xcolor}

\def\mw{{microwave}}
\newcommand{\gf}{}
\newcommand{\al}{\bf }

\begin{document}

   \title{Coupling between magnetic reconnection, energy release, and particle acceleration in the X17.2 2003 October 28 solar flare}
\titlerunning{Magnetic reconnection, energy release, and particle acceleration in the X17.2 2003 October 28 solar flare}
   \author{Victoria G. Kurt\inst{1}
          \and
          Astrid M. Veronig\inst{2,3}
          \and
          Gregory D. Fleishman\inst{4,5}
          \and
            Jürgen Hinterreiter\inst{2,6}
          \and
          Johannes Tschernitz\inst{2}
           \and
          Alexandra L. Lysenko\inst{7}
          }

   \institute{Skobeltsyn Institute of Nuclear Physics, Lomonosov Moscow State University, Moscow, 119991, Russia
         \and
            Institute of Physics, University of Graz, Universitätsplatz 5, 8010 Graz, Austria
            \and
            Kanzelhöhe Observatory for Solar and Environmental Research, University of Graz, Kanzelhöhe 19, A-9521 Kanzelhöhe, Austria 
                        \and
             Institute for Solar Physics-KIS, 79104 Freiburg, Germany
            \and
            Center for Solar-Terrestrial research, Physics Dept., New Jersey Institute of Technology, Newark, NJ 07102 USA
            \and
          Institute of Space Research, Austrian Academy of Sciences, Schmiedlstraße 6, 8042 Graz, Austria
            \and
            Ioffe Institute, St. Petersburg, 194021,  Russia
             }

   \date{}

 
  \abstract
   {The 2003 October 28 (X17.2) eruptive flare was a unique event. The coronal electric field and the $\pi$-decay $\gamma$-ray emission flux had the highest values ever inferred in solar flares.}
   {Our aim is to reveal  physical links between the magnetic reconnection process, the energy release, and the acceleration of electrons and ions to high energies in the chain of the magnetic energy transformations in the impulsive phase of the solar flare.}
   {The global reconnection rate $\dot{\varphi}\left(t\right)$ and the local reconnection rate (coronal electric field strength) $E_{\mathrm{c}} (\vec{r},t)$ are calculated from flare ribbon separation in H$\alpha$ filtergrams and photospheric magnetic field maps. HXRs measured by CORONAS-F/SPR-N, and the derivative of the GOES SXR flux, $\dot{I}_{\mathrm{SXR}}(t)$, are used as proxies of the flare energy release evolution. The flare early rise phase, main raise phase, and main energy release phase were defined based on temporal profiles of the above proxies. Available results of INTEGRAL and CORONAS-F/SONG observations are combined with Konus-Wind data to quantify the  time behavior of electron and proton acceleration.
   Prompt 
   $\gamma$-ray lines and delayed 2.2 MeV line temporal profiles observed with Konus-Wind and INTEGRAL/SPI used to detect and quantify the nuclei with energies of 10–70 MeV. 
   }
   {The magnetic-reconnection rates $\dot{\varphi}\left(t\right)$ and $E_{\mathrm{c}} (\vec{r},t)$ follow a common evolutionary pattern with the proxies of the flare energy released into high-energy electrons.
   The global and local reconnection rates reach their peaks at the end of the main rise phase of the flare. The spectral analysis of the high-energy $\gamma$-ray emission revealed a close association between the acceleration process efficiency and the reconnection rates. High-energy bremsstrahlung continuum and narrow $\gamma$-ray lines were observed in the main rise phase when $E_{\mathrm{c}} (\vec{r},t)$ of the positive (negative) polarity reached values of $\sim$120$\ \mathrm{V\ cm^{-1}}$ ($\sim$80$\ \mathrm{V\ cm^{-1}}$). 
   In the main energy release phase, the upper energy of the bremsstrahlung spectrum was significantly reduced and the pion-decay $\gamma$-ray emission appeared abruptly.
   We discuss the reasons why the change of the acceleration regime occurred along with the large‐scale magnetic field restructuration of this flare.
   }
  {The similarities between the proxies of the flare energy release with $\dot{\varphi}\left(t\right)$ and $E_{\rm c} (\vec{r},t)$ 
  in the flare main rise phase are in accordance with the reconnection models. We argue that the main energy release and proton acceleration up to subrelativistic energies began just when the reconnection rate was going through the maximum i.e., after a major change of the flare topology.}


   \maketitle

\section{Introduction}

Solar flares are energetic transient phenomena that involve several phases of energy transformation. After a complex preflare build up of the free magnetic energy in the solar corona, an instability leads to a sudden release of the free energy during a flare by magnetic reconnection {\gf driven by strong, likely turbulent, plasma flows \citep[e.g.,][]{Chen2020} and accompanied by a strong dynamic electric field \citep{Fleishman2020}}. The tracers of this released energy are the emission observed from radio to gamma-rays, solar energetic particles (SEPs), and coronal mass ejections (CME).
A significant portion of the energy  is spent on the acceleration of charged particles and plasma heating{\gf , which includes heating due to the collisional energy loss of the flare-accelerated particles} \citep[see, for example,][]{Martens1990,Litvinenko1995,Veronig2005,Holman2016, Fleishman2020, Fleishman2022} {\gf and direct heating \citep{Gary1989,2010ApJ...725L.161C,Caspi2014,Fl_etal_2015}}; {\gf see also} the reviews by \citet{Fletcher2011}; \citet{Vilmer2011}; \citet{Holman2011}; \citet{Somov2013}). However, it is not yet clear how exactly the magnetic energy release leads to the efficient acceleration of high-energy particles. The fundamental force capable of accelerating a charged particle is due to an electric field \citep[see][]{Jokipii1979}. This electric field in flares can be created by a variety of processes, such as a voltage drop at the reconnection site, MHD turbulence, or different kinds of MHD shock waves that are produced during flares \citep[e.g., review by][]{Miller1997}. 
These accelerated high-energy electrons and protons produce various components of the nonthermal microwave, X-ray, and 
$\gamma$-ray spectrum \citep{Bastian1998,Dermer1986,Ramaty1987,Murphy1987}, while the thermal plasma heated during the flare produces enhanced emission in the extreme ultra-violet (EUV) and soft X-ray (SXR) wavelength ranges.

Magnetic reconnection is an intrinsically three-dimensional (3D) process, and so, for the last twenty years, one of the main foci of researchers has been on the structure and dynamics of 3D reconnection \citep[e.g.,][]{Li2021}. Nevertheless, the quasi-2D, so called “standard flare model” \citep[CSHKP,][]{Carmichael1964,Sturrock1966,Hirayama1974,Kopp1976}, which is a 2.5D configuration with translation symmetry, successfully explains the morphology of some eruptive flares, quasi-parallel chromospheric ribbons and their divergent motion in the course of a flare. Within the CSHKP framework, a magnetic flux system can become unstable and at some point rise to higher coronal altitudes. Below it, a current sheet develops, towards which the ambient antiparallel magnetic field is brought in close contact and forced to reconnect \citep{Priest2000,Forbes2000,Vrsnak2016}. Ribbons can thus be regarded as tracers of the low-atmosphere footpoints (FPs) of newly reconnected coronal magnetic field since they are magnetically connected to the instantaneous coronal reconnection site permitting the accelerated particles to precipitate to the chromosphere. As the reconnection region moves upwards, field lines anchored at progressively larger distances from the polarity inversion line (PIL) are swept into the current sheet and reconnect. Thus, the ribbons appear farther away from the PIL as the flare-loop system is growing, leading to an apparent expansion motion of the H$\alpha$ flare ribbons \citep{Fletcher2011}. 
Measuring the expansion of the flare ribbons/FPs mapped onto the underlying magnetic field and the “shear” angle $\vartheta$ (made by the line connecting the conjugate flare FPs and the line perpendicular to the PIL), we can get insight into the magnetic reconnection process.


{\gf Regions} near/above the top of the flaring loops {\gf were proposed as possible acceleration sites after  \citet{Masuda1994} report of a coronal above-the-looptop HXR source in a flare with partly occulted fooptponts \citep{1995ApJ...444L.115W}. There are other reports on the detection of above-the-looptop acceleration sites based on X-ray and/or \mw\ data}
\citep{Sui2004,Krucker2010,Fleishman2011,Liu2013,Fleishman2020,Chen2021} {\gf including the direct mapping of the acceleration region performed using \mw\ imaging spectroscopy data \citep{Fleishman2022}} that agrees with the morphology implied by the standard flare model. It is often assumed that this is also the site of the proton (and ion) acceleration during the impulsive flare phase. Electron beams precipitate down to 
create isolated HXR FPs, brightenings, and ribbons in the ultraviolet (UV) 
and in optical chromospheric spectral lines, most prominently in H$\alpha$, with conjugate flare FPs/ribbons located on opposite sides of the magnetic polarity inversion line \citep[e.g.,][]{Veronig2006,Fletcher2011,Krucker2011,Su2013} also in agreement with the standard model. 

Models of particle acceleration and transport in solar flares suggest that magnetic reconnection and turbulence play an important role. For example, \citet{Litvinenko1995} and \citet{Petrosian2012} proposed that subrelativistic proton production begins in the turbulent reconnecting current sheet. 
This is consistent with the observational findings in
\citet{Warren2018} who reported \ion{Fe}{xxiv} (192.04~\AA) line broadening in the plasma sheet of the eruptive 2017 September 10 flare located at the solar limb and observed edge-on. The observed line width results from a combination of thermal and nonthermal broadening, favoring a strong non-thermal turbulent velocity broadening with approximately 70--150~km~s$^{-1}$ in the plasma sheet.
Recently, several solar flare studies modeled particle acceleration in a single large-scale reconnection layer and in flare termination shocks \citep[e.g.,][]{Kontar2017,Kong2022,Li2022} where turbulent energy can play a key role in the energy transfer.

Accelerated electrons can carry a large fraction of the total energy released during a flare \citep{Krucker2010,Fleishman2011,Aschwanden2017,Fleishman2022}. 
Electron beams are an important source of energy and momentum that drive the response of the low atmosphere, which determines many observable characteristics of the flare \citep[e.g.,][]{Kontar2011,Benz2017}. 
Accelerated electrons produce a non-thermal hard X-ray (HXR) and $\gamma$-ray bremsstrahlung continuum that extends up to the highest energy of the electrons themselves \citep{Miller1989,Ramaty1993}. 
The HXR intensity emitted in bremsstrahlung is proportional to the flux 
of the nonthermal 
electrons, which is linked  
to the flare energy release rate \citep[e.g.][]{Wu1986,Hudson1991}. 
The thermal soft X-ray (SXR) emission due to bremsstrahlung (free–free and free–bound emission) from the heated plasma includes a response to the energy input by the accelerated electrons in the lower atmosphere. 
Often, the time derivative of the SXR intensity, $\dot{I}_{SXR}\left(t\right)=dI_{SXR}/dt$, is proportional to the time variation of the nonthermal emission, which can be the microwave or hard X-ray emissions. 
This is known as the Neupert effect \citep{Neupert1968}. 
In practice, nonthermal bremsstrahlung with photon energies of  $>$25 keV and $\dot{I}_{SXR}\left(t\right)$ are often used as a proxy of the evolution of the flare energy conversion rate during the flare impulsive phase \citep{Hudson1991,Dennis1993,Veronig2002,Veronig2005,Dennis2003}. 

Another energetically important component of flares is accelerated ions \citep{Emslie2012}. In contrast to nonthermal electrons, which produce HXR emission only by a single process (bremsstrahlung), accelerated protons/ions contribute via several distinct emission processes observable in the $\gamma$-ray domain above roughly 500~keV. The importance of the $\gamma$-rays with photon energies $\ge$500~keV as a probe of energetic ions accelerated in solar flares was pointed out by \citet{Lingenfelter1967} and \citet{Lingenfelter1969}, who calculated the expected $\gamma$-ray fluxes. Flare-accelerated protons with energies of tens of MeV excite nuclei of the ambient matter, which emit de-excitation prompt $\gamma$-ray lines in the 0.5–12 MeV energy range \citep{Ramaty1987,Murphy2007,Murphy2009}. The strongest lines are at 4.4~MeV (\element[][12]{C}*) and 6.1~MeV (\element[][16]{O}*). Neutron capture on ambient hydrogen in the photosphere produces a strong delayed narrow line at 2.223~MeV. There are numerous less intense lines and a continuum of unresolved line emission. Protons with energies in the $\sim$70--200~MeV range in the solar atmosphere are practically undetectable, since there are no nuclear reactions with a threshold in this energy range resulting in prominent $\gamma$-ray radiation \citep{Murphy2007,Murphy2009}. If the protons gain energies above $\sim$200~MeV, they  produce neutral and charged pions through interactions with the ambient solar nuclei (the threshold of the p-$\alpha$ reactions is about 200~MeV, of p-p is close to 300~MeV; these  processes are dominant due to the elemental abundances in the corona). These pions decay and generate $\gamma$-ray emission with a specific spectrum, with a broad plateau in the 30–150 MeV energy range \citep[e.g.][]{Murphy1984,Ramaty1987,Murphy1987}. The pion-decay ($\pi$-decay) $\gamma$-ray emission during a solar eruptive event traces proton acceleration up to high energies (>200~MeV) and their interaction with the dense layers of the solar atmosphere. When high-energy protons interact with the matter, the $\pi$-decay $\gamma$-rays are emitted almost instantaneously. Thus, the solar flare gamma-ray spectrum is a superposition of bremsstrahlung due to electrons and positrons, $\gamma$-ray lines, and $\pi$-decay emissions due to nuclei.

In this paper, we investigate in detail the impulsive phase of the 2003 October 28 eruptive flare in order to reveal links between the evolution of the energy release quantified by the magnetic reconnection rates and the acceleration of electrons and nuclei to high energies. The paper is structured in the following way. 
Section~\ref{sec:magrec} covers the basics of the experimental estimation of magnetic reconnection characteristics.
A brief overview of the 2003 October 28 flare is presented in Section \ref{sec:overview}. Section~\ref{sec:data_analysis} describes the data and data analysis. 
In Section~\ref{subsec:assoc_rec_hep} we bring the results of the previous sections together and compare the evolution of populations of high-energy accelerated particles with the time evolution of $\dot{\varphi}\left(t\right)$ and $E_{\mathrm{c}} (\vec{r},t)$. 
In Sect. \ref{sec:summary}, we discuss the results and present our conclusions.

\section{Basics of magnetic reconnection rates}
\label{sec:magrec}

Magnetic-reconnection theories predict how fast reconnection can occur. Since the work by \citet{Forbes1984}, observations of chromospheric H$\alpha$/EUV flare ribbons and kernels as well as HXR FPs together with photospheric magnetic fields were used to estimate the magnetic reconnection rates and reconnection electric fields in terms of “global” and “local” reconnection rates in the low corona. \citet{Forbes1984} derived a simple relationship between the coronal electric field $E_c (\vec{r},t)$, which represents a “local” reconnection rate, and the apparent motion of the chromospheric flare ribbons, which exists in a two-dimensional configuration with translational symmetry along the third dimension. The coronal electric field $E_c (\vec{r},t)$ in two-ribbon flares can be derived as the cross product of two observables---the apparent flare-ribbon separation speed $\vec{v}_{\perp}$ and the chromospheric magnetic field strength component $\vec{B}_{n}$ perpendicular to the solar surface:
\begin{equation}
    E_{c}\left(\vec{r},t\right)=\frac{1}{c}\vec{v}_{\perp}\times\vec{B}_{n} 
    \label{eq:evxb}
\end{equation}
where $c$ is the speed of light. Given that both $\vec{v}_{\perp}$ and $\vec{B}_{n}$  are perpendicular to the ribbon, the electric field $E_c (\vec{r},t)$, is parallel to the ribbon. In practice, the following assumptions are often adopted: $\vec{B}_{n}$ does not change significantly from the photosphere to the chromosphere, and does not change during the flare. For active regions (ARs) near the disk center, $\vec{B}_{n}$ can be approximated by the line-of-sight (LOS) component of the photospheric magnetic field in pre-flare magnetograms.
We note, however, that during strong events, the photospheric magnetic field component $\vec{B}_{n}$ in flare ribbon can show flare-related changes \citep[reviews by][]{Toriumi2019,Petrie2019}.

Equation \ref{eq:evxb} has been applied to a number of flares \citep[e.g.,][]{Qiu2002,Qiu2005,Asai2004,Krucker2005,Miklenic2007,Temmer2007,Fleishman2016,Fleishman2020,Hinterreiter2018}. These studies showed that in strong flares, $E_c$ can reach dozens of V cm\textsuperscript{-1}. $E_c$ peak values reported in the literature for X-class flares belong to the range of 1--80~V~cm$^{-1}$ \citep[e.g.,][]{Asai2004,Qiu2004,Jing2005,Hinterreiter2018}. \citet{Temmer2007} reported that the highest reconnection rates $E_{c}\left(\mathbf{r},t\right)$ are typically found at locations that map to HXR FPs, whereby the difference in $E_{c}\left(\mathbf{r},t\right)$ for locations with/without HXRs is about one order of magnitude. 

\citet{Forbes2000} generalized Equation \ref{eq:evxb} to overcome the limitations of the 2D magnetic field configuration. They considered the magnetic flux of one polarity $\varphi$ swept by the flare ribbons:
\begin{equation}
    \varphi=\iint_{A}\vec{B}\cdot d\vec{a}
    \label{eq:phiba}
\end{equation}
and its time derivative:
\begin{equation}
    \dot{\varphi}=\int_{C}\left(\vec{B}\times\vec{V_{R}}\right)\cdot \mathrm{d}\vec{l}=c\oint_{C}\vec{E}_{0}\cdot \mathrm{d}\vec{l},
    \label{eq:e0dl}
\end{equation}
where $\vec{E}_{0}$ is the electric ﬁeld along a PIL, $\vec{B}$ is the magnetic ﬁeld vector measured in the photosphere, and $C$ is the curve surrounding the newly closed area $A$,  d$\vec{a}$ and d$\vec{l}$ are the area element and the arc element, respectively. Equation \ref{eq:e0dl} gives the voltage drop along the PIL and corresponds to the rate of open ﬂux converted to closed ﬂux. Several statistical studies derived the global and local reconnection rates $\dot{\varphi}$ and $E_c$, from UV or H$\alpha$ flare ribbon observations \citep{Jing2005,Qiu2005,Kazachenko2017,Tschernitz2018,Hinterreiter2018}, demonstrating correlations between the flare class, CME speed and the reconnection rates. \citet{Tschernitz2018} found a high correlation ($r = 0.9$) of the peak values of the global reconnection rate with the SXR peak flux and its derivative over 4 orders of magnitude in GOES flare class.

A correspondence has been also noted between the time evolution of $\dot{\varphi}\left(t\right)$ and the time profiles of SXR derivative or HXR emission in the flare impulsive phase \citep[e.g.,][]{Qiu2009,Qiu2010,Miklenic2009,Veronig2015}.
However, \citet{Miklenic2009} found that in several eruptive flares the peak of the reconnection rate occurred $\sim$1--2 minutes earlier than the main HXR peak. \cite{Naus2022} report a delay of three minutes of the HXR peak relative to the reconnection rate peak for the M7.3 flare of 2014 April 18. 
A similar time difference was also found between the maximum of the SXR derivative and the reconnection rate peaks 
\citep{Qiu2009,Yushkov2023}.

\section{Overview of 2003 October 28 flare characteristics}
\label{sec:overview}

Retrospective analysis shows that the largest solar ARs host photospheric magnetic field densities up to 6 kG  \citep{2006SoPh..239...41L}, coronal magnetic fields up to 4 kG \citep{2023ApJ...943..160F}, and contain free magnetic energy of $\sim$$10^{34}$~ergs sufficient to produce “superflares” \citep{Toriumi2017}. The eruptive X17.2 flare of 2003 October 28 (recalibrated by \cite{Hudson2023} as X25.7)  occurred in AR 10486 which contained a large amount of free magnetic energy of $\ge$$6\cdot 10^{33}$~ergs \citep[see, e.g.,][]{Veselovsky2004,Metcalf2005}. It was one of the most powerful flares observed during the space era. Strong emissions were observed in meterwave, microwave, submillimeter, optical, ultraviolet, SXR, HXR and $\gamma$-ray wavelengths during the flare impulsive phase. The associated CME was very fast with a speed of about 2500~km~s\textsuperscript{-1} \footnote{\url{http://cdaw.gsfc.nasa.gov/CME_list}}. High-energy solar neutrons were observed 
by SONG \citep{Kuznetsov2011} and by the near-equatorial (vertical cut-off rigidity = 9.12 GeV) Neutron Monitor (NM) Tsumeb, Namibia \citep{Bieber2005, Plainaki2005}. Also, solar energetic particles and a ground-level enhancement (GLE 65\footnote{\url{http://www01.nmdb.eu/nest}}) were associated with this flare. 

\begin{table*}[htb]
    \centering
    \caption{Overview of 2003 October 28 flare parameters.}
\begin{tabular}{lll}
\hline\hline
N  & Parameter                                                                                                        & Value              \\
\hline
1  & SXR/H$\alpha$ importance                                                                                         & X17.2/4B\tablefootmark{a}           \\
2  & {\gf Peak value of} $\dot{I}_{\mathrm{SXR}}$ (W m\textsuperscript{-2} s\textsuperscript{-1})                                         & $6\cdot 10^{-6}$\tablefootmark{b}    \\
3  & Thermal energy of the SXR-emitting plasma, (erg)                                                                 & $>19\cdot 10^{30}$\tablefootmark{c}  \\
4  & Total reconnected flux $\varphi$ (Mx)                                                                               & $(14.8\pm 2.7)\cdot 10^{21}$ \tablefootmark{b}*;  $(17.3\pm 2.1)\cdot 10^{21}$ \tablefootmark{d}                  \\
5  & Magnetic flux change rate $\dot{\varphi}$, (Mx s\textsuperscript{-1})                                                &  $(4.4\pm 0.9)\cdot 10^{19}$\tablefootmark{b};  3.4$\cdot 10^{19}$\tablefootmark{e}         \\
6  & $E_{c}$, (V cm\textsuperscript{-1})                                                                              &  80-120 \tablefootmark{b} ; $68.1\pm 3.4$ \tablefootmark{f} ; 40 \tablefootmark{g}                    \\
7  & Bremsstrahlung flux at 300 keV, (phot cm\textsuperscript{-2} s\textsuperscript{-1} MeV\textsuperscript{-1})      &  $5.5\cdot 10^2$ \tablefootmark{b}                   \\
8  & Bremsstrahlung flux at 20 MeV, (phot cm\textsuperscript{-2} s\textsuperscript{-1} MeV\textsuperscript{-1})       &  $4.2\cdot 10^{-3}$ \tablefootmark{i}                   \\
9  & $\gamma$-ray line flux at 4.4+6+1 MeV (phot cm\textsuperscript{-2} s\textsuperscript{-1})                        &  0.67 
\tablefootmark{h}                   \\
10 & $\pi$-decay emission flux at 100 MeV (phot cm\textsuperscript{-2} s\textsuperscript{-1} MeV\textsuperscript{-1}) &  $1.1\cdot 10^{-2}$ \tablefootmark{i}        \\        
\hline
\end{tabular}
\tablefoot{Numbers in lines 1, 2, 5-10 present peak values \\
* The total reconnected flux $\varphi$ was calculated during the impulsive phase until 11:13:30 UT 
}
\tablebib{ (a) Solar Geophysical Data\footnote{\url{https://www.ngdc.noaa.gov/stp/solar/sgd.html}}; (b) present work; (c) \citet{Emslie2012}; (d) \citet{Qiu2005}; (e) \citet{Tschernitz2018}; (f) \citet{Hinterreiter2018}; (g) \citet{Liu2009}; (h) \citet{Kiener2006}; (i) \citet{Kuznetsov2011}.
}
    \label{tab:characteristics}
\end{table*}

In Table \ref{tab:characteristics}, we summarize several important characteristics of this flare.
The majority of the outlined quantities  are exceptional. The amount of magnetic flux participating in the reconnection process estimated as $\sim$$2\times 10^{22}$~Mx indicates that this flare was a very powerful event. The estimated local electric field was very high and in some locations equals to $\sim$100--120~V~cm$^{-1}$ that is favorable to accelerate particles to high energies \citep{Jokipii1979}. The 2003 October 28 flare produced
the most powerful 
$\pi$-decay $\gamma$-ray emission to date with the flux reaching $1.1\times 10^{-2}$~phot~cm\textsuperscript{-2}~s\textsuperscript{-1}~MeV\textsuperscript{-1} at the photon energy of 100 MeV \citep{Kuznetsov2011}.

\section{Data Analysis and Results}
\label{sec:data_analysis}

\subsection{Data Sources}
\label{subsec:data_sources}

We used H$\alpha$ filtergrams from Udaipur Solar Observatory\footnote{\url{https://www.prl.res.in/prl-eng/division/usob}}, LOS magnetograms from the Michelson Doppler Imager \citep[MDI;][]{Scherrer1995}, full-Sun GOES X-ray fluxes,  CORONAS-F/SPRN data in two energy bands, 15--40 keV and 40--100 keV \citep{Zhitnik2006}, and  $\gamma$-ray observations from Konus-Wind\footnote{\url{ http://www.ioffe.ru/LEA/kwsun/}} \citep{Aptekar1995, Lysenko2022}. 
Our analysis is supplemented with a suitable set of observations: time profiles of HXR
and $\gamma$-ray lines taken by the $\gamma$-ray spectrometer SPI onboard the INTEGRAL spacecraft in the energy range from 600 keV to 8 MeV and by the Anti-Coincidence Shield of SPI above 150 keV \citep{Kiener2006}, as well as some results from 
previous studies which have used HXR and $\gamma$-ray images by RHESSI \citep{Hurford2006},
solar radio flux data from the USAF Radio Solar Telescope Network (RSTN)\footnote{\url{https://www.ngdc.noaa.gov/stp/space-weather/solar-data/solar-features/solar-radio/rstn-1-second/}}   and EUV loop analysis from TRACE by \citet{Su2006}. 
We used the CORONAS-F/SONG observations of the flare $\gamma$-ray emission and  spectral analysis results in the energy interval 1--150\,MeV reported by \cite{Kuznetsov2011}. 


\subsection{Global reconnection rates }
\label{subsec:global_rec_rate}

To derive the global, $\dot{\varphi}\left(t\right)$, and local, $E_\mathrm{c}\left(\vec{r},t\right)$, reconnection rates, the following data preprocessing was used (for details, we refer to \cite{Tschernitz2018}). For the analysis, we used an USO H$\alpha$ filtergram sequence available at starting 5 min before the flare. The time cadence of the USO H$\alpha$ observations is about 30 s. All images were rotated to solar north up and then differentially rotated to the time of the first image of the H$\alpha$ image sequence. For intensity normalization we applied to each image a zero-mean and whitening transformation described in \citet{Poetzi2015}. A sub-region containing the flare region was selected, and all H$\alpha$ images of the observing sequence were co-aligned with the first image of the series using spatial cross-correlation techniques to compute and account for the x- and y- shifts in the images due to seeing. A pre-flare MDI LOS magnetogram (pixel size of 2.0 arcsec) was selected closest in time to the first H$\alpha$ image, co-aligned and regridded to the pixel size of the USO H$\alpha$ images (0.6 arcsec).

\begin{figure*}
    \centering   \includegraphics[width=0.8\textwidth]{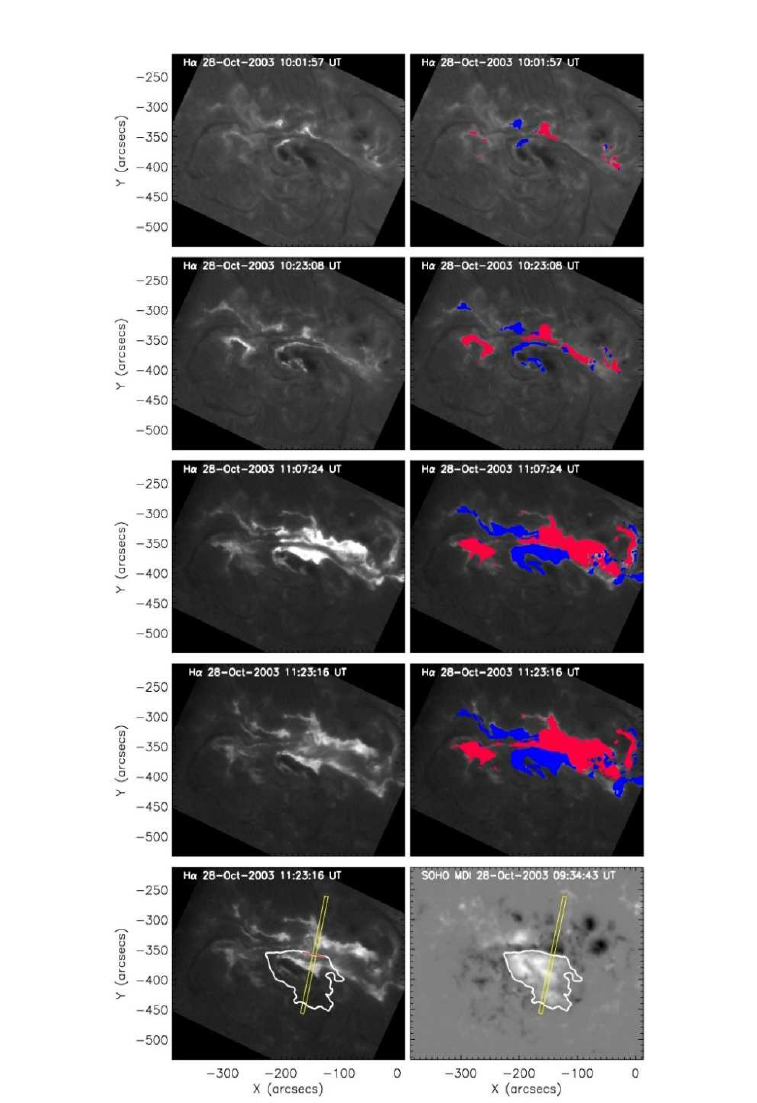}
    \caption{Evolution of the 2003 October 28 flare observed by the USO H$\alpha$ telescope for four different time steps covering the start, maximum and decay phase of the flare (left panels). In the right panels, the total flare ribbon areas detected up to the respective recording time of the image are overplotted, separately in blue/red for the positive/negative magnetic polarities. In the bottom panels, we outline the flare PIL (white contour) derived from the corresponding pre-flare MDI LOS magnetic map shown in the right bottom panel. The narrow yellow rectangle in the bottom panels represents an example of one direction perpendicular to the local PIL segment (red line) along which the flare ribbons separation motion and $E_{\mathrm{c}}\left(\vec{r},t\right)$ were derived.}
    \label{fig:flare_ev}
\end{figure*}

To calculate the “global” reconnection rates $\dot{\varphi}\left(t\right)$, we identified for each time step of the H$\alpha$ image sequence the newly brightened flare pixels with respect to the previous time steps. To identify flaring pixels, we used a thresholding technique based on increases of >5.5 standard deviations in the intensity-normalized H$\alpha$ images. Finally, for each time step we applied the derived mask of newly brightened flare pixels to the LOS magnetic field maps, and calculated the magnetic flux swept by newly brightened flare pixels to obtain for each time step $t$ \citep[for further details of the calculations see][]{Veronig2015,Tschernitz2018}. 

Figure \ref{fig:flare_ev} gives an overview of the 28 October 2003 flare evolution in USO H$\alpha$ images along with the derived cumulated flare pixel masks (blue/red for positive/negative magnetic polarity) used to calculate $\dot{\varphi}\left(t\right)$. The bottom panels show the corresponding MDI LOS magnetic field map and illustrate the ribbon evolution normal to the local PIL for the  $E_c$ calculation.

\begin{figure}
  \includegraphics[width=0.49\textwidth]{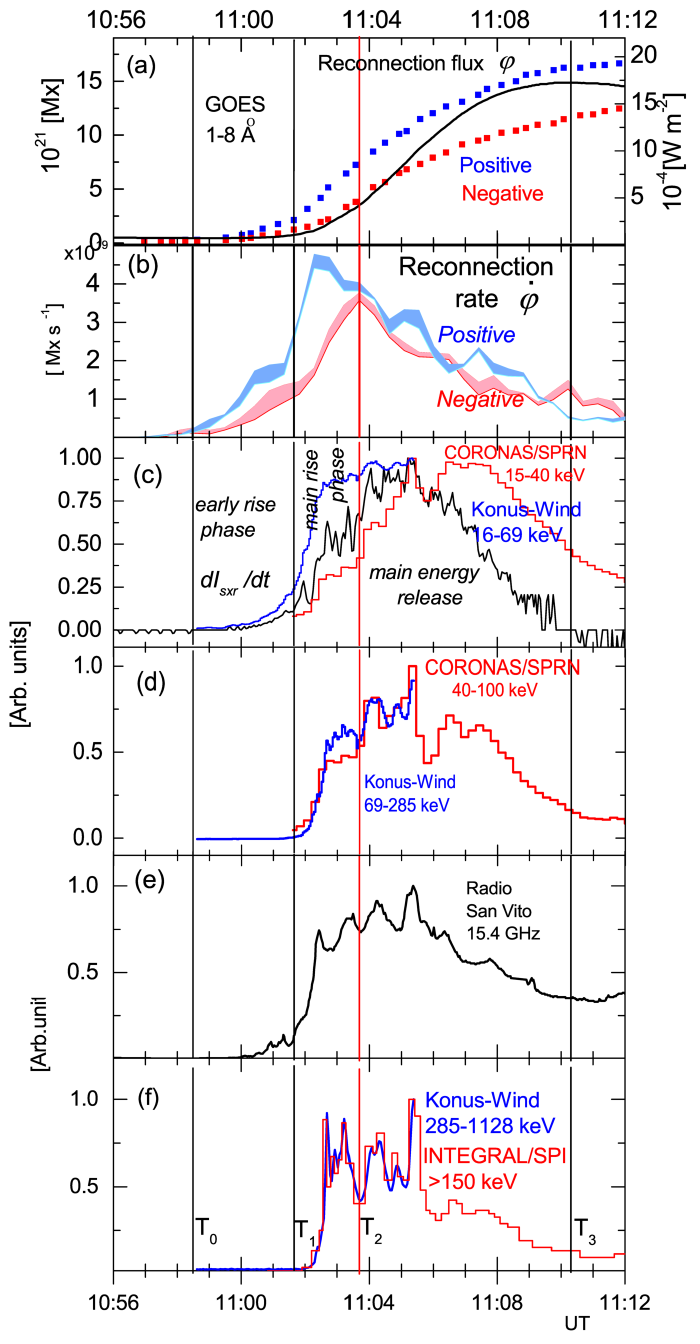}
    \caption{Observables of the 2003 October 28 flare. (a) Reconnection flux derived in the positive (blue) and negative (red) domains (left Y-axis) and the GOES 1--8~\AA\ soft X-ray flux $I_{SXR}\left(t\right)$ (black curve, right Y-axis). (b) Reconnection rates $\dot{\varphi}\left(t\right)$ positive(blue) and negative(red). (c) $\mathrm{d}I_{SXR}/\mathrm{d}t$ (black), Konus-Wind count rate at 16--69\,keV (blue) and CORONAS-F/SPRN rate at 15--40\,keV (red). (d) Konus-Wind rate at 69--285 keV (blue) and CORONAS-F/SPRN rate at 40--100\,keV(red). (e) 15.4\,GHz microwave emission flux (black). (f) INTEGRAL/SPI flux above 150\,keV (from \cite{Kiener2006}, red) and Konus-Wind rate at 285--1128\,keV(blue).  The early rise phase onset time $T_0$=10:58:30~UT and the main rise phase onset time $T_1$=11:01:16~UT are marked with black vertical lines. The red vertical line indicates the onset of the main energy release phase, $T_2$=11:03:40~UT, that ended at about $T_3$=11:10~UT. }
    \label{fig:manifestations}
\end{figure}

Figure \ref{fig:manifestations} plots the GOES 1--8~{\AA} SXR flux along with the inferred time evolution of the reconnection fluxes derived during the flare impulsive phase, separately for the positive and negative magnetic fields. The reconnection flux $\varphi\left(t_{k}\right)$ is defined as the sum of all fluxes in the flare areas that brightened up until a given time $t_{k}$. Ideally, $\varphi_{+}$ and $\varphi_{-}$ should be identical, as equal amounts of positive and negative magnetic flux participate in the reconnection. However, measurements do not always yield a perfect balance between the positive and negative fluxes \citep[e.g.,][]{Fletcher2001}. Given the uncertainties involved in the measurements, flares with $R = \varphi_{+}\left(t\right)/\varphi_{-}\left(t\right)$ from 0.5 to 2.0 are generally regarded as showing a good flux balance \citep{Qiu2005,Miklenic2009,Tschernitz2018}. Figure \ref{fig:manifestations}(a) shows that our measurements obey this requirement.

The total flare reconnection flux $\varphi\left(t\right)$ is defined as the mean of the absolute values of the reconnection fluxes in both polarity regions at the end of the time series, i.e.
\begin{equation}\varphi\left(t\right)=\left(\vert\varphi_{+}\left(t\right)\vert + \vert\varphi_{-}\left(t\right)\vert\right)/2
    \label{eq:tot_rec_flux}
\end{equation}
where $\varphi_{+}\left(t\right)$ and $\varphi_{-}\left(t\right)$ are the cumulated magnetic fluxes in the positive and negative polarity regions.  The total amount of magnetic flux participating in the reconnection up to the time 11:13:30 UT was $(14.8\pm 2.7)\cdot 10^{21}$~Mx. 
The magnetic-flux change rates $\vert\dot{\varphi}_{+}\left(t\right)\vert$ and $\vert\dot{\varphi_{-}}\left(t\right)\vert$ were calculated as the time derivative of the  reconnected flux separately for each magnetic polarity. 
These rates are presented in Figure \ref{fig:manifestations}(b). 
{\gf The mismatch between these two values is  largest during the rise phases (before the red line in Figure\,\ref{fig:manifestations}), which might be associated with a greater inclination of the field in this early phase of the flare consistent with the lower altitude
inferred for the X-point in Section\,\ref{subsec:loc_e_field}.
 }
$\vert\dot{\varphi}_{+}\left(t\right)\vert$ reached its absolute maximum equal to $\left(4.6\pm 0.2\right)\cdot 10^{19}$~Mx~s\textsuperscript{-1} at 11:02:35 UT. $\vert\dot{\varphi_{-}}\left(t\right)\vert$ reached maximum at 11:03:41~UT with a lower value $\left(3.7\pm 0.1\right)\cdot 10^{19}$~Mx~s\textsuperscript{-1}. Afterward, the reconnection rates declined but were still at a level $\geq 1\cdot 10^{19}$~Mx~s\textsuperscript{-1} for at least 10 minutes, implying ongoing magnetic reconnection. 

\subsection{Energy release and high energy emissions }
\label{subsec:en_release}

Figure \ref{fig:manifestations}(c) shows the temporal profile of the time derivative of the GOES 1--8~{\AA} flux, $\dot{I}_{SXR}\left(t\right)$, and the HXR emission at 15--40\,keV registered by CORONAS-F/SPR-N and at 16--69\,keV registered by Konus-Wind. 
Figure \ref{fig:manifestations}(d) shows time profiles of the nonthermal HXR emission at 40--100 keV from CORONAS-F/SPR-N and at 69--285 keV from Konus-Wind. 
CORONAS-F/SPR-N data are useful because the SPR-N detector had a small geometric factor and therefore it was not saturated during this intense flare. Figure \ref{fig:manifestations}(e) shows 15.4 GHz radio-emission from the RSTN.
These data demonstrate that the Neupert effect \citep{Neupert1968,Veronig2002} revealed itself in the flare under consideration. Therefore, we will use hereinafter the GOES SXR derivative, $\dot{I}_{SXR}\left(t\right)$, as a proxy of the flare energy release. Figure \ref{fig:manifestations}(f) shows HXR emission 
observed by the Anti-Coincidence Shield of SPI above 150 keV \cite{Kiener2006} (red) and Konus-Wind rate at 285--1128 keV (blue).

Konus-Wind detected 16--69 keV HXR emission at the flare beginning simultaneously with the 
15.4 GHz radio emission indicative of electron  acceleration up to $\gtrsim$100 keV when the reconnection rates 
$\dot{\varphi}\left(t\right)$ of both polarities have not yet exceeded $(1-2)\cdot 
10^{18}$~Mx~s\textsuperscript{-1}. 
We adopted the time $T_0$=10:58:30 UT as the flare onset and divided the flare impulsive phase based on the $\dot{\varphi}\left(t\right)_{mean}$, $I_{SXR}$, $\dot{I}_{SXR}$ and HXR time profiles as follows: the early rise phase: 10:58:30–11:01:16 UT ($T_0$–$T_1$), the main rise phase: 11:01:16–11:03:40 UT ($T_1$–$T_2$), the main energy release phase: 11:03:40–11:10:00 UT ($T_2$–$T_3$). We adopted the time $T_3$ as the end of the main energy release phase. Thus, the main energy release phase began and developed when the global reconnection rate has already gone over the maximum.

\begin{figure*}
    \centering   
    \includegraphics[width=0.8\textwidth]{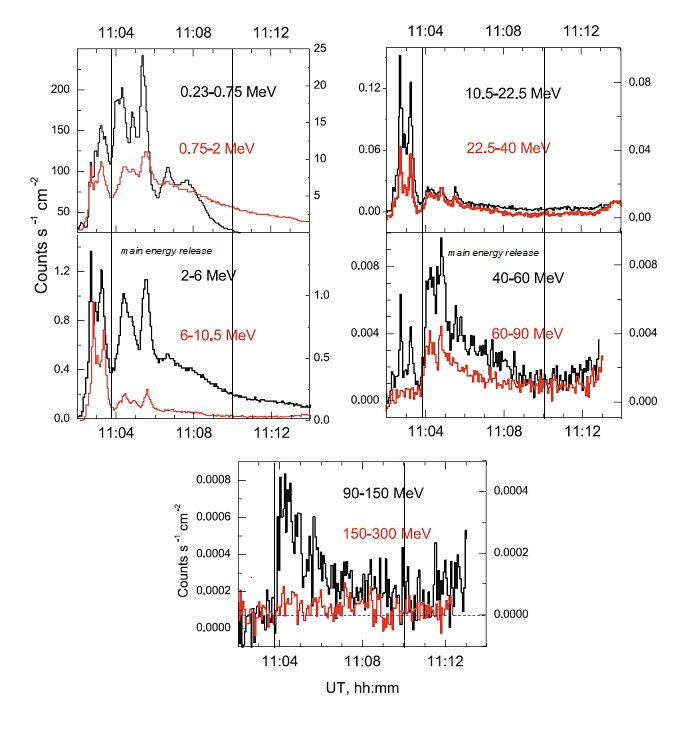}
    \caption{Time evolution of CORONAS/SONG response during the flare of 28 October 2003 starting at 11:02:00~UT. The background count rates are subtracted. The left axis in each panel corresponds to the lower energy channel, the right axis to the higher energy one. Reproduced from Figure 5 of \cite{Kuznetsov2011} under Springer Nature License Number \# 5605391357254. 
    }
    \label{fig:SONG_LCs}
\end{figure*}

Figure~\ref{fig:SONG_LCs} reproduced from  Fig. 5 of \cite{Kuznetsov2011} presents the time profiles of high-energy emission measured by CORONAS-F/SONG in the wide energy channels from 225 keV up to 150 MeV. 
SONG observations of the flare began after CORONAS-F has left the outer radiation belt at 11:02~UT and covered most part of the main rise phase and the entire main energy release phase. 
Two vertical lines added in Figure~\ref{fig:SONG_LCs} indicate the beginning and the end of the main energy release phase: 11:03:40--11:10:00~UT ($T_2$--$T_3$). 
Figures \ref{fig:manifestations} and  \ref{fig:SONG_LCs} show  that particle acceleration/energy release happened as several separate episodes. As it is expected from the standard flare model, {\gf \sout{each significant episode of particle acceleration occurred simultaneously with an episode of}} the reconnection rate {\gf variations agree with those of particle acceleration \sout{changes}}. Indeed, such a {\gf \sout{coincidence} correspondence} being in agreement with the model of \cite{Vrsnak2016} is  seen in both the early rise and in the main rise phases. However, the main energy release began later, around $\sim$11:03:45~UT when the positive polarity reconnection rate 
reached its peak 
and the negative polarity one has already passed the maximum. The main energy release maximum was observed at $\sim$11:05:20$\pm$00:00:06~UT.

\subsection{Local electric field strength}
\label{subsec:loc_e_field}

\begin{figure*}
    \centering    \includegraphics[width=1.\linewidth]{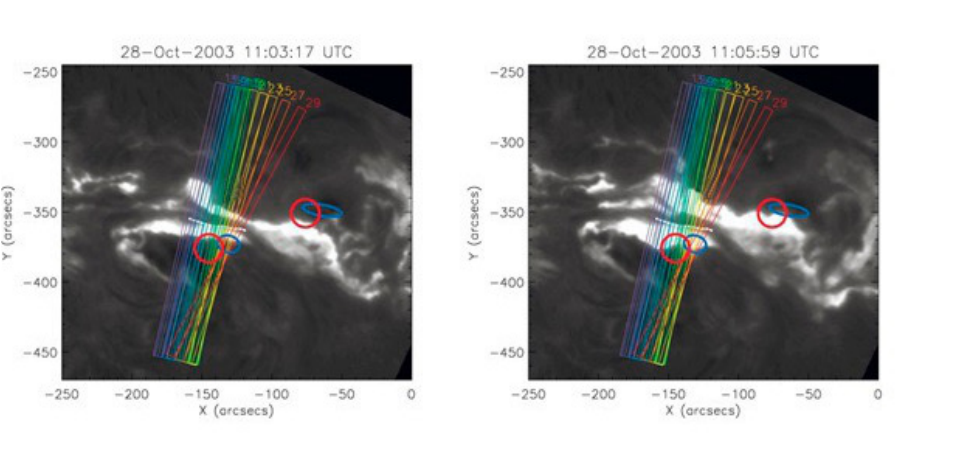}
    \caption{H$\alpha$ images observed at the main rise phase (left) and at the main energy release phase peak (right). The local PIL segment is shown by the white line. Multicolor rectangles indicate the directions perpendicular to the local PIL segments, along which we followed the flare ribbon separation to derive the local electric field strength evolution $E_{c}\left(\vec{r},t\right)$. Overlaid on the H$\alpha$ image is the RHESSI HXR 100--200~keV (blue) and the 2.2~MeV $\gamma$-ray line (red) contours for 11:06:20-11:09:20 UT  \citep[from][]{Hurford2006}. The two HXR FPs lie on the outer borders of the extended conjugated H$\alpha$ ribbons. We note that one of the two footpoints (rooted in negative polarity) is located outside the 30 positions along the PIL that we are studying.}
    \label{fig:loc_rec_dir}
\end{figure*}

To study the evolution of the local reconnection rates (coronal electric fields) $E_{c}\left(\mathbf{r},t\right)$, we manually selected positions along the PIL, from where we track the flare ribbon separation motion. The PIL location was selected in such a way that the conjugate flare ribbons can be tracked simultaneously in a direction perpendicular to the PIL. The analysis is performed for 30 positions along the PIL and the different colors correspond to the selected tracking directions (see Figure \ref{fig:loc_rec_dir}). Each rectangle perpendicular to the PIL indicates the sub-region (length of $200''$ and width of $6''$) used to track the ribbons. In the tracking, the outer front of the flare ribbons needs to be identified, because this part is related to the newly reconnected field lines along which the accelerated particles travel downwards to the solar surface. For a detailed description of the method and the uncertainties we refer to \citet{Hinterreiter2018}. We determine $E_{c}\left(\vec{r},t\right)$ by tracking the ribbon position as a function of time and evaluating the underlying photospheric magnetic field strength. Then the $E_{c}\left(\vec{r},t\right)$ value in each location was calculated using Eq. \ref{eq:evxb}. 

In Figure \ref{fig:loc_rec_dir} we overplot also contours of the 100–200~keV bremsstrahlung and the 2.2~MeV $\gamma$-ray line emission sources reconstructed from RHESSI observations in \citet{Hurford2006}. These images were obtained after the spacecraft came out of the Earth shadow at 11:06:43 UT. Two footpoints (FPs) are observed in HXRs (electron bremsstrahlung) positioned on localized regions at the outer borders of the extended conjugate H$\alpha$ ribbons. We restricted the analysis up to the time 11:08:00 UT when considering the $E_{c}\left(\vec{r},t\right)$ time evolution. 

\begin{figure}
    \centering
    \includegraphics[width=0.5\textwidth]{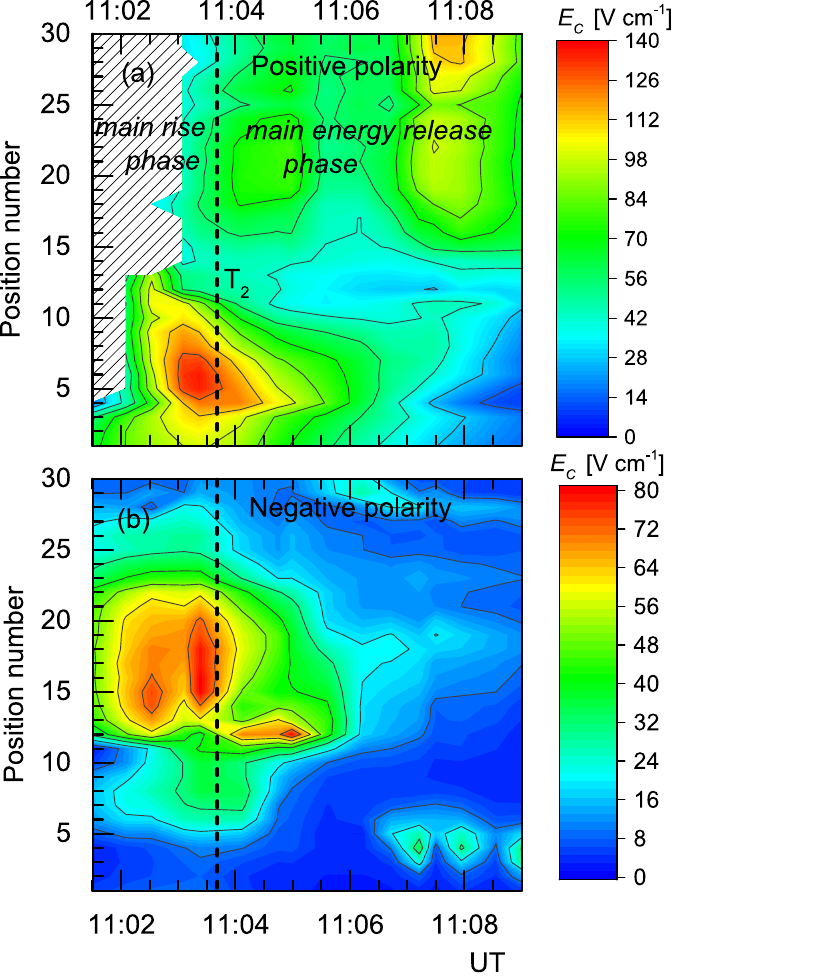}
    \caption{Spatio-temporal map of the coronal electric field strengths $E_{c}\left(\vec{r},t\right)$. (a) Positive polarity contour plot. The shaded area marks the region in which the electric field could not be determined; (b) Negative polarity contour plot.  The dashed vertical line indicates hereinafter $T_2$=11:03:40 UT.}
    \label{fig:ec_map}
\end{figure}

\begin{figure}
    \centering
    \includegraphics[width=0.5\textwidth]{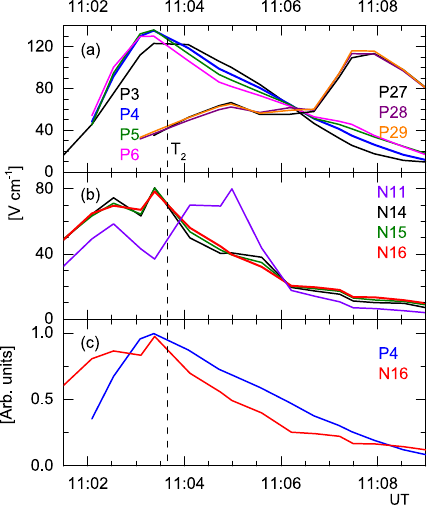}
    \caption{Evolution of the coronal electric field $E_{c}\left(\vec{r},t\right)$ in selected positions of positive (a) and negative (b) magnetic polarities. (c) Comparison of the $E_{c}\left(\vec{r},t\right)$ evolution in positions P4 and N16 (cf. Fig. \ref{fig:loc_rec_dir}).}
    \label{fig:ec_evo}
\end{figure}

Figure \ref{fig:ec_map} shows contour plots of $E_{c}\left(\vec{r},t\right)$ at each position number vs time in the positive (P) and negative (N) polarities derived for 30 positions during the flare impulsive phase. The evolution of the $E_{c}\left(\vec{r},t\right)$  at the different ribbon locations reflects the propagation of magnetic reconnection along the arcade of the conjugate loops rooted in positive and negative polarities. The $E_{c}\left(\vec{r},t\right)$ distributions look slightly different in the two polarities. This is because a slit in a given direction across the local PIL used to derive the $E_{c}$ profiles, does not necessarily cover the conjugated FPs in both polarities, because the overall flare loop (and ribbon) system is highly sheared. We found that:
\begin{itemize}
    \item In the positive polarity domain, the highest electric field strength $E_{c,max}\approx$ 125\,V\,cm$^{-1}$. Values $\geq$95\% of $E_{c,max}$ were observed between 11:03:10 and 11:03:40~UT in position numbers P4–P6. 
    \item In the negative polarity domain, we obtain $E_{c,max} \approx80$\,V\,cm$^{-1}$, and  values $\geq$95\% of $E_{c,max}$ were observed at position numbers N14–N16 at the same time as for the positive polarity.
    \item $E_{c}\left(\mathbf{r},t\right)$ of both polarities starts to fall off around 11:03:20–11:03:40 UT, i.e. around $T_2$ at the main flare energy release onset.
    \item Afterwards, the peak  $E_{c}\left(\vec{r},t\right)$ of the  negative polarity was observed around N11 position  with a 1 minute delay relative to $T_2$ (see Figure \ref{fig:ec_evo}(b)). A similar peak in the positive polarity was less pronounced as it merged with the main peak.
\end{itemize}

The $E_{c}\left(\mathbf{r},t\right)$ evolution at the P4–P6 and N15–N16 positions are shown in Figures \ref{fig:ec_evo}(a),(b). According to the standard model of solar flares, a pair of flare FPs in the chromosphere maps the reconnected magnetic field lines in the current sheet, whose apex runs close to the “X-point”. We adopted positions P4 and N16 as most representative ones. The Pearson correlation coefficient between $E_{c}\left(\vec{r},t\right)$ in these positions is $r = 0.95$, indicative of a magnetic connectivity between them. 

\begin{figure*}
    \centering
    \includegraphics[width=0.99\textwidth]{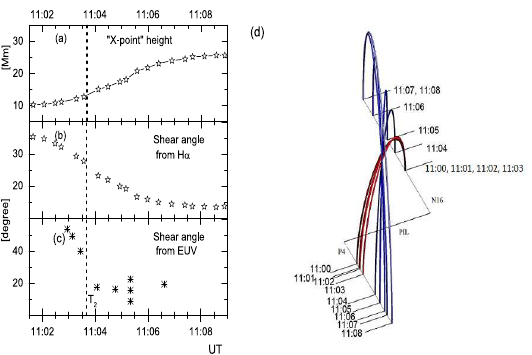}
    \caption{(a) Evolution of the height of the reconnection X-point derived from the loop connecting positions P4 and N16. (b) Shear angle derived from the$ E_c$ evolution of positions P4 and N16. (c) Shear angle derived from EUV observations \citep[from Figure 10 of][]{Su2006}. 
    (d) Axonometric projection cartoon of the time evolution of the presumable loop connecting positions P4 and N16 (side view).}
    \label{fig:heigt_xpoint}
\end{figure*}

We calculated the distance between the outer fronts of the H$\alpha$ ribbons located in the P4 and N16 positions and the shear angle between them. The P4 position crosses the PIL at $x = -156''$, $y = -357''$, the N16 position crosses the PIL at $x = -140''$,  $y= -360''$. Assuming a semicircle shape of the loop connecting these FPs, we estimated the height of the reconnection X-point at each time step (Figure \ref{fig:heigt_xpoint}(a)). A cartoon of the time evolution of this loop seen at a side view is presented in Figure \ref{fig:heigt_xpoint}(d). The derived height of the X-point was $\sim$8–12~Mm in the main rise flare phase showings a slow rise (v$\sim$20 km s\textsuperscript{-1}). Around $T_2$, it began to ascend faster ($v \sim 60$~km~s\textsuperscript{1}). Figure \ref{fig:heigt_xpoint}(b) shows the evolution of the shear angle derived from P4 and N16 positions, whereas Figure \ref{fig:heigt_xpoint}(c) shows the evolution of the shear angle derived from the observations of the EUV strongest brightening pairs \citep[from Figure 10 of][]{Su2006}. It is seen that the shear angles decrease during the main rise phase. The change of the velocity of X-point ascent and the change in shear angle occurred roughly simultaneously around 11:03–11:04 UT, i.e. close to $T_2$=11:03:40 UT. 

\subsection{High-energy X-ray and $\gamma$-ray emission spectra}
\label{subsec:high_en_spectra}

In this Section we present and discuss the observations of the flare high-energy emission, the results of Konus-Wind spectral analysis along with published INTEGRAL/SPI and CORONAS/SONG spectra.

\subsubsection{The early flare phase and the main rise phase}
\label{subsubsec:high_en_spectra_early_main}

To specify the $\gamma$-ray emission components and the spectral characteristics, 
we used Konus-Wind data.
A detailed description of the Konus-Wind data, response files and spectra deconvolution can be found in \cite{Lysenko2022}. Energy spectra were accumulated in the energy range $\sim$0.35–15\,MeV (60 channels).
The accumulation of multichannel spectra starts with the Konus-Wind trigger time 11:01:12~UT (corrected for the light propagation to the Earth) and ended at 11:02:57~UT. 
The $\gamma$-ray spectra were fitted using the XSPEC package \citep{Arnaud1996} with the following components.
The 
continuum 
was described by the sum of two components: (i) the power-law (PL), and (ii) the power-law with an exponential cutoff at higher energies (CPL). 
De-excitation $\gamma$-ray lines were fitted with the template proposed by \cite{Murphy2009}.
Electron-positron annihilation and neutron-capture lines were fitted by gaussians with fixed positions at 511\,keV and 2223\,keV and fixed $\sigma$ at 5\,keV and 0.1\,keV, respectively. 

\begin{figure*}
    \centering
    \includegraphics[width=0.33\linewidth]{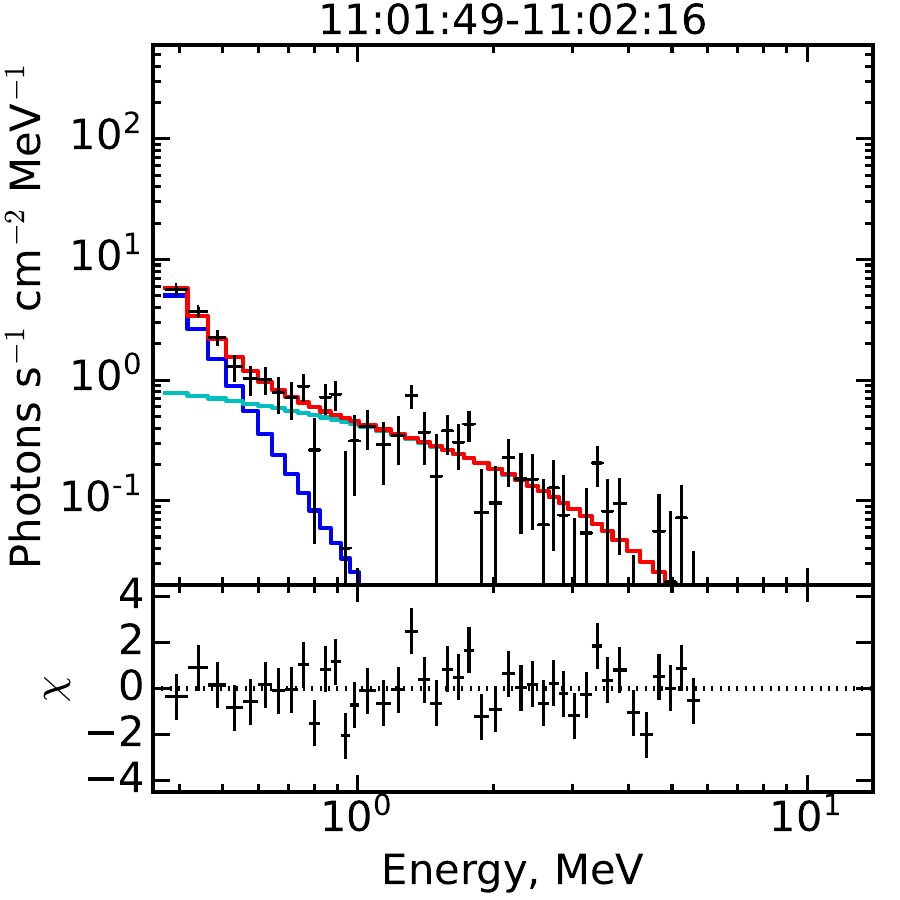}
    \includegraphics[width=0.33\linewidth]{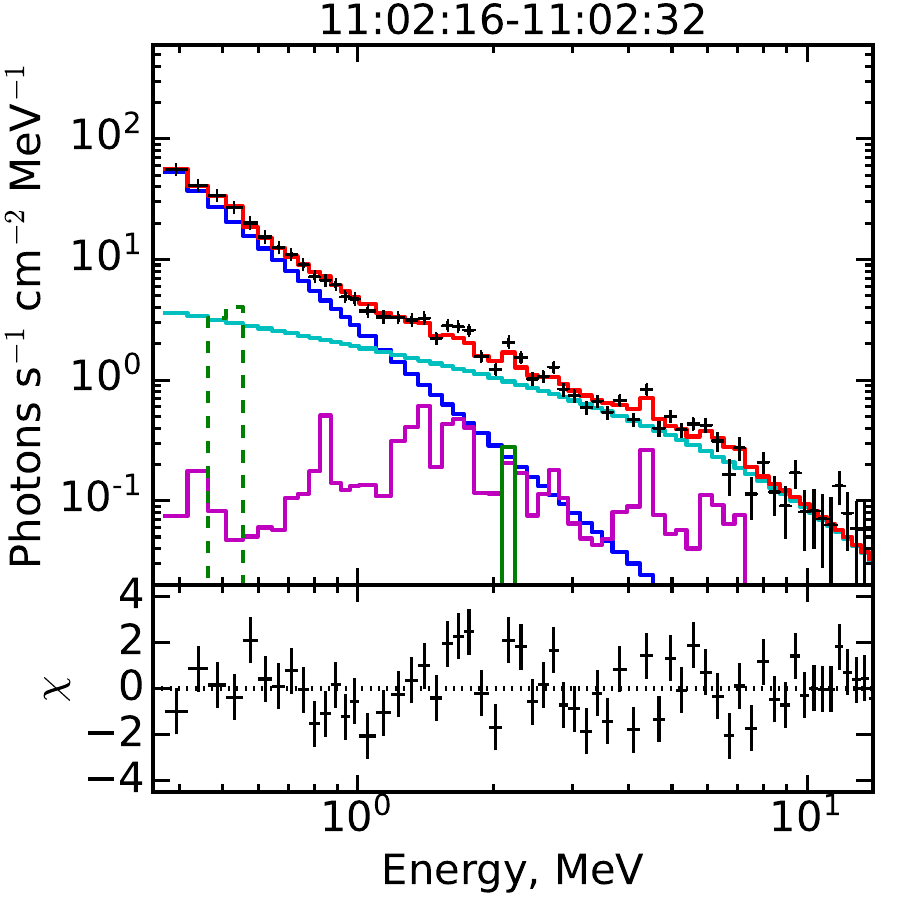} 
    \includegraphics[width=0.33\linewidth]{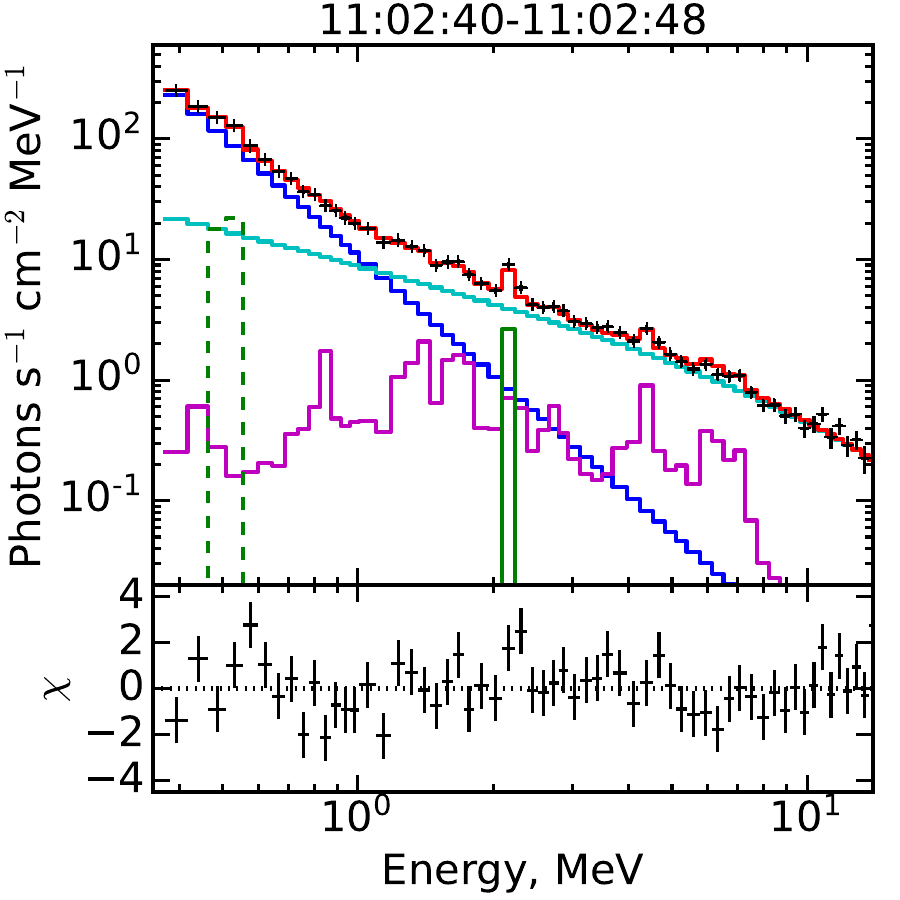}
    \caption{Konus-Wind spectra accumulated over three time intervals during the main rise phase: 11:01:49--11:02:16 (left), 11:02:16--11:02:32 (middle), 11:02:40--11:02:48 (right). Black crosses represent the photon spectrum, color curves indicate model components: PL (blue), CPL (cian), nuclear deexcitation lines (magenta), neutron capture line (green), electron-positron annihilation line (dashed green), and the sum of all components (red). The bottom panels on each plot represent the fit residuals.} 
    \label{fig:kw_spectra}
\end{figure*}

In Figure~\ref{fig:kw_spectra}, we present Konus-Wind spectra accumulated subsequently during the main rise phase of the flare. 
\footnote{{\al At the time of the flare, Konus-Wind was $\sim$3.25 light seconds farther from the Sun than the Earth, thus, for the comparison to the CORANAS/SONG results fluxes obtained by Konus-Wind have to be multiplied by a factor of $\sim$1.014. 
As this value lies within uncertanties we did not correct photon fluxes in Figure~\ref{fig:kw_spectra} to account for the distance difference.}} 
This spectral analysis shows: 
\begin{itemize}
    \item A line-free, very soft electron bremsstrahlung component consistent with the PL model with a spectral index of $\sim$5 during the time interval 11:01:12--11:01:49\,UT (not shown in the Figure).
    \item The appearance of emission above 1\,MeV at 11:01:49~UT, related to the hard continuum component described by the CPL model (Figure~\ref{fig:kw_spectra}, left).
    \item The appearance of $\gamma$-ray lines, including de-excitation lines, electron-positron annihilation line, and neutron capture line resulting from nuclear interactions of accelerated ions in the solar atmosphere at 11:02:16~UT (Figure~\ref{fig:kw_spectra}, middle). 
    \item The CPL continuum exceeds 14\,MeV during time interval 11:02:40--11:02:48~UT (Figure~\ref{fig:kw_spectra} right), which coincides with the time of the first maximum of CORONAS-F/SONG observations in the 40--60\,MeV energy range at 11:02:36-11:02:48~UT (see Figure\,\ref{fig:SONG_LCs}).
    \item $\gamma$-ray lines are barely seen above the continuum. This is consistent with the results of the high-resolution $\gamma$-ray spectrometer INTEGRAL/SPI. Due to the exceptional intensity of bremsstrahlung, INTEGRAL/SPI distinguished only two narrow $\gamma$-ray lines in the  600\,keV--8\,MeV energy range, namely at 2.223\,MeV and 4.4\,MeV.
\end{itemize}

\begin{figure}
    \centering
    \includegraphics[width=0.49\textwidth]{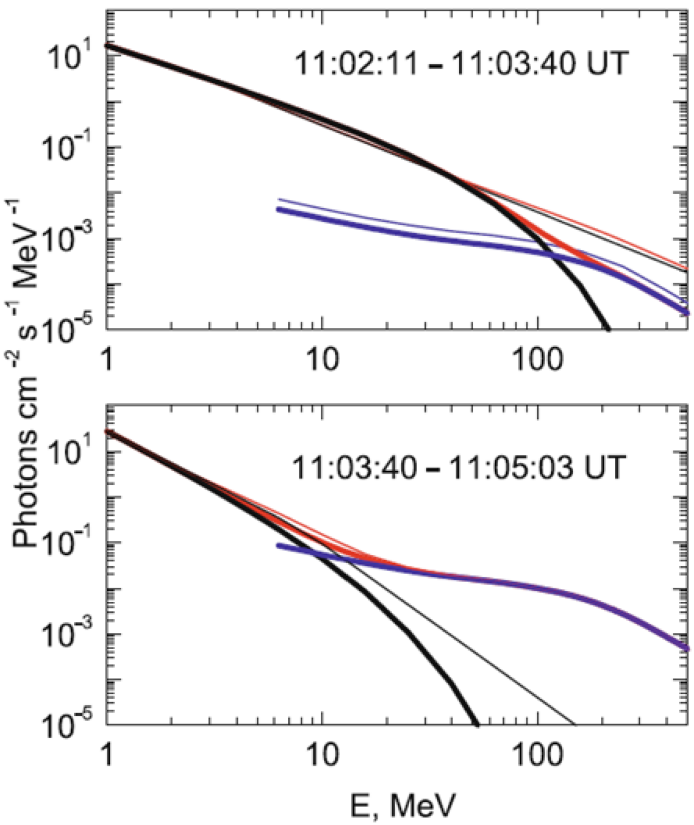}
    \caption{Gamma-ray emission spectra observed by CORONAS/SONG; from Figure 8 of \cite{Kuznetsov2011}, reproduced under Springer Nature License Number \# 5605391357254. Upper panel: spectrum accumulated over the main rise phase. Bottom panel:spectrum accumulated over the most part of the main energy release phase. The red curves represent the total spectra, the black curves indicate the continuum component (thin curves correspond to a power law with varying index and the thick curves correspond to a power law with an exponential cutoff), while the blue curves show the $\pi$-decay component.}
     \label{fig:SONG_spectra}
\end{figure}

In the following, we put these findings into context with the SONG spectral analysis in \cite{Kuznetsov2011}. 
SONG had 12 broad energy channels and did not resolve discrete narrow $\gamma$-ray lines including the neutron-capture line at 2.223 MeV. The top panel of  Figure\,\ref{fig:SONG_spectra} shows the spectrum accumulated over the main rise phase; the bottom panel shows the spectrum accumulated in the main energy release phase.
\cite{Kuznetsov2011} adopted a two-component model: an electron bremsstrahlung component described by a power-law above 1\,MeV with an index gradually changing with energy or as an exponential high-energy cutoff; the second component, $\pi$-decay emission, having a broad ‘line’ due to neutral-pion decay, peaking at 67 MeV, plus a continuum due to bremsstrahlung from electrons and positrons produced in charged pion decay \citep{Murphy1987}. \cite{Kuznetsov2011} emphasized that they did not explicitly take into account the contribution of the $\gamma$-ray lines, as they have focused on the $\pi$-decay component; this would overestimate the continuum if a significant $\gamma$-ray line contribution were present. 
\cite{Kuznetsov2011} also performed a similar fit using channels above 10 MeV, i.e., explicitly avoiding those lines. The results of both fittings showed that neglecting the contribution of $\gamma$-ray lines did not change the resulting $\pi$-decay flux. 
The fits of Konus-Wind spectra obtained with higher spectral resolution confirm that the $\gamma$-ray line contribution were weak, which justifies its omission by \cite{Kuznetsov2011}. Therefore, the SONG fit showed that the primary electrons were accelerated at least up to 22.5--40\,MeV, and that the estimated upper limit of the $\pi$-decay emission is $\leq5\times10^{-4}$\,phot\,cm$^{-2}$\,s$^{-1}$\,MeV$^{-1}$ at 100\,MeV. 

\subsubsection{The main energy release phase}
\label{subsubsec:high_en_spectra_main}

This flare phase lasted  $\sim$6 min (11:03:40–11:10:00 UT) (see Figures \ref{fig:manifestations}, \ref{fig:SONG_LCs}). The accumulation of Konus multichannel spectra terminated at 11:02:57~UT. We used results of INTEGRAL/SPI and GORONAS-F/SONG spectral analysis of this phase. \cite{Kiener2006} found clear evidence of the bremsstrahlung continuum softening and the increase of the $\gamma$-ray line intensity in this phase (see their Table~2). 
The SONG $\gamma$-ray spectrum in the time interval between 11:03:40--11:05:03~UT (bottom panel of Figure~\ref{fig:SONG_spectra}) shows that the continuum became negligible compared to the $\pi$-decay component at energies $>$22.5\,MeV. 
An important point is that the count rate of the high energy SONG channels, $>$90\,MeV, (Figure~\ref{fig:SONG_LCs}) which obviously was due to the $\pi$-decay emission, had emerged around the main flare energy release beginning at $T_2$=11:03:40~UT. 
\cite{Kuznetsov2011} followed the time evolution of the $\pi$-decay emission until 11:10:20--11:12:20\,UT. This component reached its maximum of $(1.1\pm0.1)\times10^{-2}$ phot cm$^{-2}$ s$^{-1}$ MeV$^{-1}$ at 11:03:40--11:05:03~UT (see the bottom panel of Figure~\ref{fig:SONG_spectra}), simultaneously with the maximum of the flare energy release. 

\subsubsection{Constraints on the accelerated ion spectrum}
\label{subsubsec:high_en_var}

The spectral hardness of the accelerated ions can be found through comparison of fluxes or fluences of different components of the $\gamma$-ray emission: the neutron-capture line 2.223 MeV, F2.2, the prompt 
4.4 MeV 12 C* and 6.1 MeV 16 O* nuclear de-excitation lines, and the pion-decay component, $F_{\pi}$. The pion production cross section has a very small value near the threshold of this reaction of 200--300 MeV, and increases by a factor of 100 when the energy of accelerated protons doubles. Thus, for hard spectra, the relevant ion energy range extends from a few MeV nucl$^{-1}$ up to $>$100 MeV\,nucl$^{-1}$. For soft spectra, the relevant ion energy range is much narrower, from less than 1 to a few MeV\,nucl$^{-1}$ (see Murphy et al. 2007). 

The primary accelerated ions are assumed to have a power-law energy distribution with spectral index $S$. \cite{Tatischeff2005} deduced the spectral index $S$ for this event from the 2.223 MeV to the 4.4 and 6.1\,MeV line fluence ratios from the INTEGRAL/SPI data in the range $3 \leq S \leq 4$. \cite{Share2004} found $S$ close to 3.4 for this flare after 11:06 UT from RHESSI data. \cite{Kuznetsov2011} combined the CORONAS/SONG $F_{\pi}$ fluxes with the nuclear deexcitation line flux in the 4--7\,MeV range, $F_{4-7}$, obtained from INTEGRAL/SPI data by \cite{Kiener2006} and derived the hardest spectrum with $S=2.8\pm0.1$ near the peak of the $\pi$-decay emission and $S=3.1\pm0.1$ afterward. In summary, these findings show that the power law index of the proton spectrum is $S\sim3$ over a wide energy range from several MeV to several hundred MeV. In comparison with other strong flares: \cite{Omodei2018} found the hardest value $S =3.2\pm0.1$  during the main energy release phase of the X8.2 2017 September 10 flare (at 15:57:55--15:58:54 UT).

\section{Association of the reconnection rates with the high-energy particle acceleration }
\label{subsec:assoc_rec_hep}


\begin{figure}
    \centering
    \includegraphics[width=0.4\textwidth]{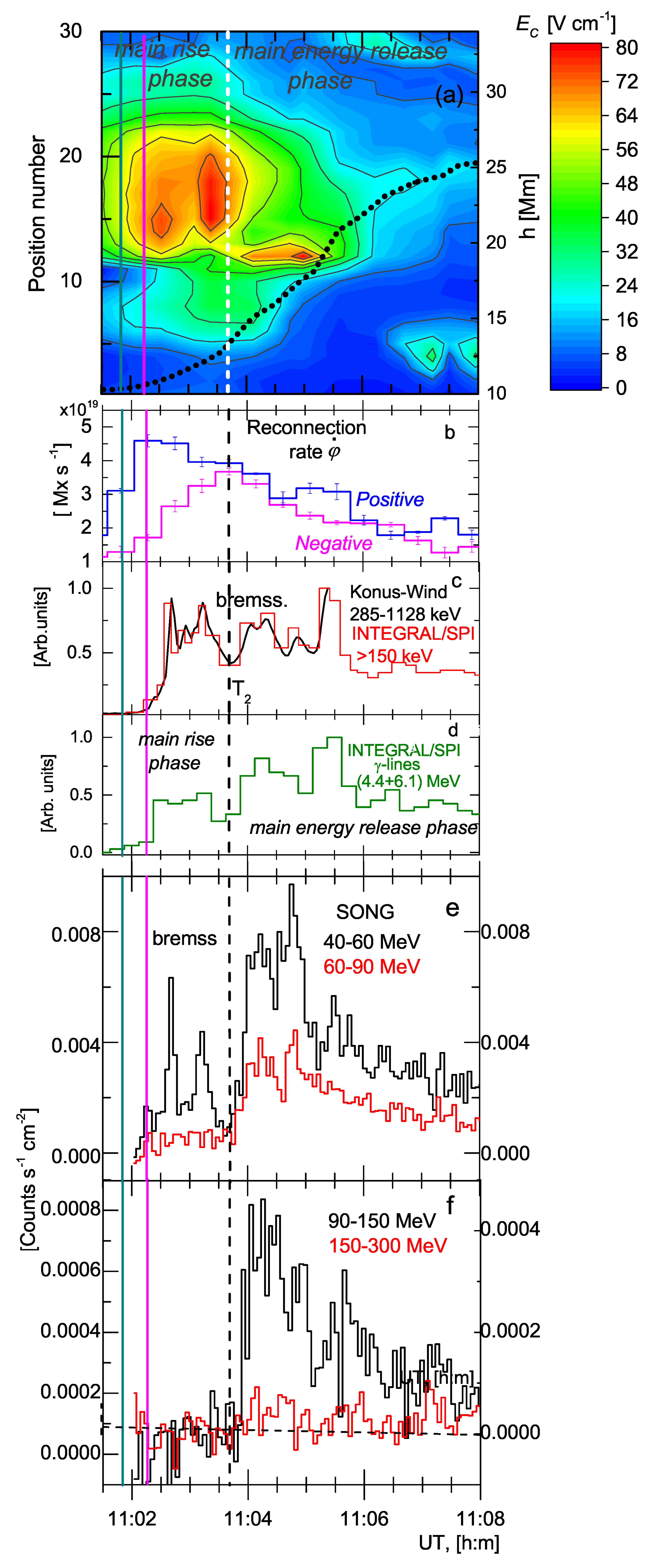}
    \caption{Local 
    and global reconnection rates  collated with high-energy electron and proton emissions during the main rise phase and the main energy release phase of the 2003 October 28 flare.  (a) Contour plot of the local reconnection rate $E_c$(t) in the negative polarity domain, overplotted with the derived height-time evolution of the X-point  (black points, right axis). (b) Time profiles of the global reconnection rates $|\dot \varphi|$, separately for the two polarities. (c) Electron bremsstrahlung at intermediate energies observed by Konus-Wind (black) and INTEGRAL/SPI (red); from \cite{Kiener2006}. (d) Prompt $\gamma$-ray lines from INTEGRAL data \citep{Kiener2006}. (e,f) Electron high-energy bremsstrahlung and $\pi$-decay emission observed by CORONAS/SONG; reproduced under Springer Nature License Number \# 5605391357254 (see also Figure \ref{fig:SONG_LCs}). The dark-cyan thick line indicates the onset of the bremsstrahlung emissions with energies $>$1 MeV at 11:01:49 UT. The magenta thick line indicates the onset of the $\gamma$-ray lines at 11:02:15 UT. The black dashed line indicates the onset of the main energy release phase at T2=11:03:40 UT.}
    \label{fig:rec_acc}
\end{figure}

Figure \ref{fig:rec_acc} shows the evolution of the reconnection rates together with various manifestations of high-energy electrons and protons. Panel 
(a) shows the distribution of the negative polarity electric field $E_{c}\left(r,t\right)$ at various positions relative to the PIL over time, along with the derived increase of the X-point height (right axes). Panel (b) shows the time evolution of the global reconnection rates of the negative and positive polarity. Panel (c) shows the
bremsstrahlung 
observed by the Anti-Coincidence Shield of SPI above 150\,keV \citep{Kiener2006}, and of 285--1128\,keV observed by Konus-Wind. Panel (d) shows the prompt 4.4 and 6.1\,MeV $\gamma$-ray line profile (from \cite{Kiener2006}). Count rates of the two CORONAS-F/SONG channels at 40--60\,MeV and 60--90\,MeV are shown in panels (e, f). One can see that time $T_2$=11:03:40 UT identified as the onset of the main energy release phase (see Sect.~\ref{subsec:en_release}) demarcates two time intervals with different signatures of charged particle acceleration. 

The hard bremsstrahlung component, which is usually described by a CPL model, appeared in the flare spectrum in the time interval 11:01:49--11:02:16~UT during the main rise phase. 
At that time, the local reconnection rate in the negative polarity reached $E_c\approx$ 60\,V\,cm$^{-1}$ and the averaged global reconnection rate $\dot{\varphi}(t)\approx$ 2$\times$10$^{19}$\,Mx\,s$^{-1}$. 
Prompt $\gamma$-ray line emission appeared later at 11:02:16~UT, where the reconnection rates reached values of $E_c\approx$ 75\,V cm$^{-1}$ and $\dot{\varphi}(t)\approx$ 3$\times$10$^{19}$\,Mx\,s$^{-1}$.
In the main rise phase two maxima of the bremsstrahlung and $\gamma$-ray line emission occurred simultaneously (within the available time resolution) at 11:02:40 UT and at 11:03:12 UT. Puting this all together, it is tempting to associate the electron and proton acceleration peaks with the above two maxima of $E_c$ (see also Figure \ref{fig:ec_map}) which in turn were associated with regions of the reconstructed X-point height $h \approx$10--12 Mm. 

The time interval between 11:03:45 and 11:10:00 UT corresponds to the flare main energy release. It is important that the temporal profiles of the global reconnection rates in both polarities, $|\dot{\varphi}_-(t)|$ and $|\dot{\varphi}_+(t)|$,  decreased after $\sim$11:03:45 UT (see Figure  \ref{fig:rec_acc}(b)). The beginning of a new process of the electron acceleration was identified based on the Konus-Wind 285--1128 keV data at 11:03:45 (Figure  \ref{fig:rec_acc}(c)). A new increase of $\gamma$-ray line flux observed by the INTEGRAL/SPI with a 15\,s time sequence occurred in the time interval 11:03:42--11:03:57~UT giving a good agreement with the electron acceleration onset time. Spectral analysis of SONG data \citep{Kuznetsov2011} showed that the count rates of the SONG channels $>$60 MeV are entirely due to $\pi$-decay emission in the time interval 11:03:47--11:04:15 UT and afterward. Moreover, Figure \ref{fig:SONG_LCs} shows that statistically significant $\pi$-decay emission appeared promptly during one bin of SONG time resolution (4 s) at 11:03:49~UT $\pm$ 2\,s.

The $\pi$-decay emission jumped promptly up to 0.75~$I_{max}$ with a time lag of 9$\pm$2~s relative to the electron acceleration onset. 
Apparently, this delay provides evidence that protons need a few extra seconds to gain enough energy to create the observable number of photons. 
Earlier, we have defined 
the main energy release phase of the flare based on the electron acceleration as a proxy (see Figure~\ref{fig:manifestations}). We have now shown that {\gf this is also a `high-energy release' phase, when} the most efficient acceleration of protons to very high energies occurred{\gf \sout{ at the same time}}. 
The height of the X-point during this energy release phase is $h \sim 15-20$~Mm. 


The new acceleration process described above, began when the contours of $E_c$ reshaped and the implied X-point has just moved up. Although there were no direct observations of the flare morphology in the corona in the 2003 October 28 flare available, as the event occurred on the solar disk, we suggest an evolving morphology similar to that reported for the well-observed 2017 September 10 X8 limb flare \citep{Gary2018, Veronig2018,Fleishman2020,Chen2020,Li2022,Kong2022}. We propose that over the course of the flare, two distinct sources of accelerated particles were sequentially formed, initially one at a height $h \sim 10-12$~Mm and then the other at $h\sim 15-20$~Mm. We attribute these regions to the cusp location and to the bottom of the reconnection current sheet. We have demonstrated that the flare main energy release and proton acceleration up to subrelativistic energies began just when the reconnection rate was going through its maximum, i.e.\ after a major change in the topology of the flare, when the erupting flux rope is already far away from the flare site.

\section{Summary and Discussion}
\label{sec:summary}


We have investigated the impulsive phase of the X17.2 eruptive flare of 2003 October 28 in order to reveal links in the chain of the evolution of the energy release as quantified by the magnetic reconnection rates and of the acceleration of electrons and protons to high energies. Our main findings are:
\begin{enumerate}
    \item The distinct phases of the energy release, namely, the rise phase and the main energy release phase, are morphologically different from each other and are also different in their associated signatures of the particle acceleration.
    \item The rise phase is associated with a larger reconnection rate (stronger coronal electric field) and relatively low positions of the implied reconnection X-point as derived from the conjugate flare ribbon locations. This phase demonstrates very efficient acceleration of nonthermal electrons and protons up to dozens of MeV, which is confirmed by detection of electron bremsstrahlung up to $\sim$40--60~MeV and the simultaneous $\gamma$-ray de-excitation lines. The relatively low position of the implied reconnection X-point is analogous to those obtained by \cite{Kurt1996, Kurt2000}, who proposed that a low height of the acceleration region could be favorable for electron acceleration up to $\sim$100 MeV.
    \item The later, main energy release flare phase is associated with a higher location of the reconnection X-point. At this stage protons are accelerated to higher energies (>200~MeV), while electrons are accelerated to lower energies ($<20$\,MeV).
\end{enumerate}
Similar relationships between the reconnection rates and particle acceleration properties at the rise and main flare phases have been noted for some other cases \citep[e.g.][]{2022AGUFMSH43A..03S, Yushkov2023} with the $\pi$-decay gamma-ray component observed in the impulsive flare phase, which can imply a fundamental character of the reported relationships for the understanding of the solar flare process.

Let us discuss why (1) at the rise phase the electrons are accelerated up to dozens (or even one hundred) of MeV, while protons not; and (2) at the following main energy release phase, the protons are accelerated to hundreds of MeV, while the electrons are accelerated only up to much smaller energies than at the rise phase. A possible reason for that could be the evolving relationships between three characteristic time scales: the acceleration rate $T_{acc}$ (the energy e-folding time scale), the energy loss rate $T_{loss}$, and the residence time $T_{res}$ of the particles in the acceleration region.
Particles can be accelerated to a given energy $E$ if the acceleration up to this energy occurs faster than the energy loss at this energy. Thus, the highest energy $E_{max}$ is defined from the equality $T_{acc}\left(E_{max}\right) = T_{loss}\left(E_{max}\right)$, provided that the energy loss time scale is shorter than the residence time. In the opposite case, the highest energy of the accelerated particle is defined by another equality, $T_{acc}\left(E_{max}\right) = T_{res}\left(E_{max}\right)$. One important difference between electrons and protons (nuclei) are their highly different mechanisms of the energy loss due to large differences in their masses. For highly relativistic electrons, the main energy loss mechanism in the magnetized coronal plasma is the synchrotron energy losses with the characteristic time scale (in seconds)
\begin{equation}
    \tau_{syn}=6\left(10^{3}~{\rm G}/B\right)^2 \left(80/ \gamma\right),
    \label{eq:syn_loss}
\end{equation}
where $B$ is the magnetic field in G and $\gamma=E/mc^2$ is the Lorentz factor; see Eq. (12.52) in \citet{Fleishman2013}. This loss is negligible for protons and other nuclei. We propose that the acceleration rate of the electrons is largest at the rise phase, when the measured reconnection rate is largest, which permits electrons to be accelerated to rather high energies of the order of 100~MeV. At this stage the flare undergoes a fast evolution, which implies that the acceleration region can evolve quickly too (moves up following the reconnection process). This means that the protons might not reside in the acceleration region long enough to be accelerated to energies higher than $\sim$100~MeV. 

Later, at the main energy release phase, the reconnection rate decreases, which can imply that the acceleration efficiency decreases proportionally. This means that the maximum energy of the accelerated electrons also decreases to obey the new balance between the acceleration and loss rates, while the electrons accelerated earlier will lose their energy over several seconds according to Eq.~\ref{eq:syn_loss} if the magnetic field in the acceleration region is about 1~kG \citep{Fleishman2020}. However, if the particles can now reside longer in the acceleration region, e.g., due to efficient particle trapping by high levels of turbulence, protons can gain much higher energies than at the rise phase. Another factor that can support ion acceleration to high energies is the decrease of the guiding magnetic field \citep{Arnold2021,Dahlin2022} due to the eruption of a flux rope, which is far away from the acceleration region at this stage of the flare. 

This reasoning is consistent with microwave imaging observations of the powerful X8 flare on 2017 September 10 that was observed on the limb, where after the eruption and the restructuring of the magnetic connectivity the acceleration site located within the cusp region revealed highly efficient trapping of the nonthermal particles at the main flare phase \citep{Fleishman2022}, presumably due to enhanced levels of turbulence. The particles reside several minutes in the acceleration region, which, in the case of protons and other ions, can permit their acceleration to rather high energies sufficient to produce the $\pi$-decay gamma-ray component.

We thank Dr. V. Bogomolov for the kindly provided CORONAS-F/SPRN data.
{We thank Kiener et al. for providing the data of INTEGRAL/SPI.}
AMV acknowledges the Austrian Science Fund (FWF) 10.55776/I4555.
This work was supported in part by NSF grant 
AGS-2121632,  
and NASA grants
80NSSC20K0718 
and 80NSSC23K0090 
to New Jersey Institute of Technology (GDF).
The work of ALL was supported by the basic funding program of the Ioffe Institute No. FFUG-2024-0002.

\bibliographystyle{aa} 
\bibpunct{(}{)}{;}{a}{}{,} 
\bibliography{references} 
%
\end{document}